\documentclass[aps,pre,showpacs,amsmath,amssymb,longbibliography]{revtex4-2}
\usepackage{graphicx}
\usepackage{xcolor}
\usepackage{float}

\newcommand{\bR}{\bold R}
\newcommand{\br}{\bold r}

\newcommand{\be}{\begin{equation}}
\newcommand{\ee}{\end{equation}}
\newcommand{\fig}[1]{Fig.~\ref{#1}}
\newcommand{\Fig}[1]{Figure~\ref{#1}}

\newcommand{\eq}[1]{Eq.~(\ref{#1})}

\graphicspath{ {./figs/} }

\begin{document}
	\title{Hidden scale invariance in the Gay-Berne model. II.\\ Smectic B phase}
	\date{\today}
	\author{Saeed Mehri}
	\affiliation{\textit{Glass and Time}, IMFUFA, Department of Science and Environment, Roskilde University, P.O. Box 260, DK-4000 Roskilde, Denmark}
	\author{Jeppe C. Dyre}\email{dyre@ruc.dk}
	\affiliation{\textit{Glass and Time}, IMFUFA, Department of Science and Environment, Roskilde University, P.O. Box 260, DK-4000 Roskilde, Denmark}	
	\author{Trond S. Ingebrigtsen}\email{trondingebrigtsen@hotmail.com}
	\affiliation{\textit{Glass and Time}, IMFUFA, Department of Science and Environment, Roskilde University, P.O. Box 260, DK-4000 Roskilde, Denmark}

\begin{abstract}
This paper complements a previous study of the isotropic and nematic phases of the Gay-Berne liquid-crystal model [Mehri \emph{et al}., Phys. Rev. E \textbf{105}, 064703 (2022)] with a study of its smectic B phase found at high density and low temperatures. We find also in this phase strong correlations between the virial and potential-energy thermal fluctuations, reflecting hidden scale invariance and implying the existence of isomorphs. The predicted approximate isomorph invariance of the physics is confirmed by simulations of the standard and orientational radial distribution functions, the mean-square displacement as a function of time, as well as the force, torque, velocity, angular velocity, and orientational time-autocorrelation functions. The regions of the Gay-Berne model that are relevant for liquid-crystal experiments can thus fully be simplified via the isomorph theory.
\end{abstract}
\maketitle

\section{Introduction}

Liquid crystals involve molecules with a high degree of shape anisotropy \cite{degennes,jurasek2017self}. This interesting state of matter is relevant in many different contexts, ranging from display applications to biological systems \cite{woltman2007liquid,de2017rod,tian2018self}. Depending on temperature and pressure, the molecular anisotropy leads to different structural phases, e.g., nematic and smectic phases with long-range orientational ordering \cite{degennes}. 

Gay-Berne (GB) models describe molecules of varying shape anisotropy spanning from elongated ellipsoids to thin disks, and GB models have become standard liquid-crystal models \cite{gay1981modification}. The GB pair potential depends on four dimensionless parameters. This is reflected in the notation GB$(\kappa, \kappa^{\prime}, \mu, \nu)$ in which the four parameters quantify the shape of the molecules and the strength of their interactions. A previous paper studied the isotropic and nematic phases of a GB model with parameters corresponding to rod-shaped elongated molecules \cite{meh22a}. It was found that this model has isomorphs in the isotropic and nematic phases, which are curves in the thermodynamic phase diagram along which the physics is approximately invariant. This paper presents a study of the same GB model in its smectic B phase, demonstrating that isomorphs exist also here.

\section{The Gay-Berne potential and simulation details}

The GB$(\kappa, \kappa^{\prime}, \mu, \nu)$ pair potential is characterized by the following four dimensionless parameters: $\kappa\equiv \sigma_e/\sigma_s$ where $\sigma_e$ and $\sigma_s$ are lengths, $\kappa'\equiv\varepsilon_{ss}/\varepsilon_{ee}$ where $\varepsilon_{ss}$ and $\varepsilon_{ee}$ are energies, and two exponents $\mu$ and $\nu$. The GB pair potential $v_{\rm GB}$ is defined as follows \cite{gay1981modification}

\begin{subequations}
	\label{GB_pot}
	\begin{align}
		v_{\rm GB}(\textbf{r}_{ij},\hat{\textbf{e}}_i,\hat{\textbf{e}}_j) &= 4\varepsilon (\hat{\textbf{r}}, \hat{\textbf{e}}_i, \hat{\textbf{e}}_j) \left[\left(\sigma_s/\rho_{ij}\right)^{12} - \left(\sigma_s/\rho_{ij}\right)^{6} \right],
		\label{GB_pot_a} \\
		\rho_{ij} &= r_{ij} - \sigma(\hat{\textbf{r}}, \hat{\textbf{e}}_i, \hat{\textbf{e}}_j) + \sigma_s\,.
		\label{GB_pot_b}
	\end{align}
\end{subequations}
Here, $r_{ij}$ is the distance between molecules $i$ and $j$, $\hat{\textbf{r}}\equiv\textbf{r}_{ij}/r_{ij}$ is the unit vector from molecule $i$ to molecule $j$, and $\hat{\textbf{e}}_i$ and $\hat{\textbf{e}}_j$ are unit vectors along the major axes of the molecules. The GB molecule mimics an ellipsoid of two diameters $\sigma_s$ and $\sigma_e$. Specifically, one defines

\begin{subequations}
	\label{GB_sigma}
	\begin{align}
		\sigma(\hat{\textbf{r}}, \hat{\textbf{e}}_i, \hat{\textbf{e}}_j) &= \sigma_s \bigg[1-\dfrac{\chi}{2} \bigg(\dfrac{(\hat{\textbf{e}}_i\cdot\hat{\textbf{r}}+\hat{\textbf{e}}_j\cdot\hat{\textbf{r}})^2}{1+\chi(\hat{\textbf{e}}_i\cdot\hat{\textbf{e}}_j)}+ \dfrac{(\hat{\textbf{e}}_i\cdot\hat{\textbf{r}}-\hat{\textbf{e}}_j\cdot\hat{\textbf{r}})^2}{1-\chi(\hat{\textbf{e}}_i\cdot\hat{\textbf{e}}_j)}\bigg) \bigg]^{-1/2}
		\label{GB_sigma_a},\\
		\chi&=\dfrac{\kappa^2-1}{\kappa^2+1}\,.
		\label{GB_sigma_b}
	\end{align}
\end{subequations}
Here $\chi$ is a shape anisotropy parameter and $\kappa$ quantifies the molecular asymmetry such that $\kappa=1$ ($\chi=0$) represents spherical molecules, $\kappa\rightarrow\infty$ ($\chi\rightarrow1$) corresponds to very long rods, and $\kappa\rightarrow 0$ ($\chi\rightarrow-1$) corresponds to very thin disks. The energy term is given by
	
\begin{subequations}
	\label{GB_epsilon}
	\begin{align}
		\varepsilon(\hat{\textbf{r}},& \hat{\textbf{e}}_i, \hat{\textbf{e}}_j) = \varepsilon_0\, 
		\left(\varepsilon_1(\hat{\textbf{e}}_i,\hat{\textbf{e}}_j)\right)^\nu
		\left(\varepsilon_2(\hat{\textbf{r}}, \hat{\textbf{e}}_i, \hat{\textbf{e}}_j)\right)^\mu
		\label{GB_epsilon_a}\\
		\intertext{in which}
		\varepsilon_1(\hat{\textbf{e}}_i,\hat{\textbf{e}}_j)&=\big(1-\chi^2(\hat{\textbf{e}}_i\cdot\hat{\textbf{e}}_j)^2\big)^{-1/2}
		\label{GB_epsilon_b}\,,\\
		\varepsilon_2(\hat{\textbf{r}}, \hat{\textbf{e}}_i, \hat{\textbf{e}}_j)&= 1-\frac{\chi'}{2}\biggl(\dfrac{(\hat{\textbf{e}}_i\cdot\hat{\textbf{r}}+\hat{\textbf{e}}_j\cdot\hat{\textbf{r}})^2}{1+\chi'(\hat{\textbf{e}}_i\cdot\hat{\textbf{e}}_j)}+ \dfrac{(\hat{\textbf{e}}_i\cdot\hat{\textbf{r}}-\hat{\textbf{e}}_j\cdot\hat{\textbf{r}})^2}{1-\chi'(\hat{\textbf{e}}_i\cdot\hat{\textbf{e}}_j)}\biggr)\,.
		\label{GB_epsilon_c}\\
		\intertext{Here}
		\chi'&=\frac{\kappa'^{1/\mu}-1}{\kappa'^{1/\mu}+1}\,
	\end{align}
\end{subequations}
is an energy anisotropy parameter. The energies $\varepsilon_{ss}$ and $\varepsilon_{ee}$ are the well depths of the potential in the side-side and end-end configurations, respectively. Unless otherwise stated, $\sigma_s$ defines the length and $\varepsilon_0$ the energy units used below. 

We simulated a system of $1372$ particles of the GB$(3,5,2,1)$ model studied previously in Ref. \onlinecite{meh22a}. The GB pair potential was cut and shifted at $r_c=4.0$ and the time step used was $\Delta t = 0.001$. The standard $NVT$ Nose-Hoover algorithm was used for the center-of-mass motion and the Fincham algorithm was used for the rotational motion \cite{fin86,fin92}. Different thermostats were applied for the translational and the rotational motions \cite{meh22a}. The molecular moment of inertia was set to unity. A home-made code for GPU computing was used; at each simulated state point 20 million time steps were taken to equilibrate the system before the production run of 67 million time steps. 

If $\bR\equiv (\br_1,...,\br_N)$ is the vector of particle coordinates and $\rho\equiv N/V$ is the particle density, the microscopic virial $W(\bR)$ is defined as $W(\bR)\equiv \partial U(\bR)/\partial\ln\rho$ in which the density is changed by a uniform scaling all particle coordinates. For an inverse power-law pair potential, $v(r)=\varepsilon (r/\sigma)^{-n}$, it is easy to see that this implies that $W(\bR)$ is a sum of pair virial contributions equal to $(n/3)v(r)$. Because the vectors $\hat{\textbf{r}}$, $\hat{\textbf{e}}_i$, and $\hat{\textbf{e}}_j$ do not change under a uniform expansion, a related result applies for the GB pair potential. Specifically, the GB pair virial is equal to $4\varepsilon (\hat{\textbf{r}}, \hat{\textbf{e}}_i, \hat{\textbf{e}}_j) [4\left(\sigma_s/\rho_{ij}\right)^{12} - 2\left(\sigma_s/\rho_{ij}\right)^{6}](r/\rho_{ij})$, and the total microscopic virial $W(\bR)$ is calculated as the sum of all pair virials. 

\begin{figure}[H]
    \centering
    \includegraphics[width=0.35\textwidth]{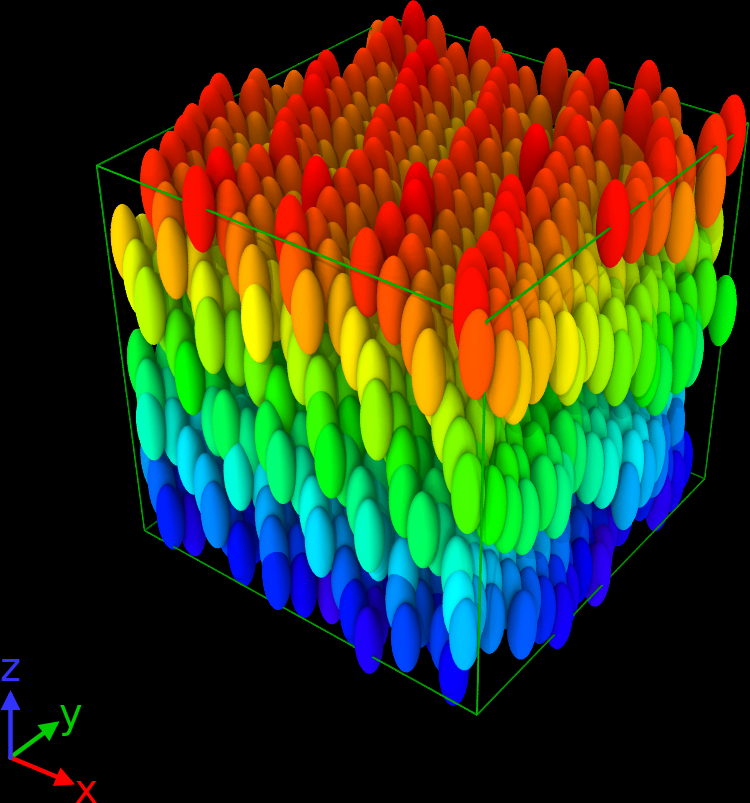}
    \caption{Snapshot of the smectic B phase at density 0.4 and temperature 1.2. A color coding is introduced here to visualize the individual planes.}
    \label{fig:rod_conf}
\end{figure}

The GB(3,5,2,1) phase diagram is shown in Fig. 3 of Ref. \onlinecite{meh22a}. Figure \ref{fig:rod_conf} shows a snapshot of the system at equilibrium in the smectic B phase.

\section{Properties studied}

The quantities evaluated numerically in this study are: the standard radial distribution function $g(r)$ \cite{bates1999computer,de1991liquid}, the below defined orientational radial distribution function $G_l(r)$ ($l=2$) \cite{ber70,de1991liquid,de1991location,adams1987computer}, and a number of single-molecule time-autocorrelation functions \cite{de1992dynamics,jose2006multiple}. The latter two observables are defined by

\begin{align}
G_l(r) \equiv \langle P_l(\hat{\mathbf{e}}_i \cdot \hat{\mathbf{e}}_j) \rangle\label{G_l},\\
	\phi_A(t)=\langle \mathbf{A}(t_0) \cdot \mathbf{A}(t_0+t) \rangle.
	\label{autocorr}
\end{align}
Here $P_l$ is the $l$'th Legendre polynomial, $\mathbf{A}(t)$ is a vector defined for each molecule, and the angular brackets denote an ensemble and particle average, which in the case of $G_l(r)$ is restricted to pairs of particles the distance $r$ apart. We study the cases of $\mathbf{A}$ being the velocity, angular velocity, force, and torque. We also study the first- and second-order molecular orientational order parameter time-autocorrelation functions defined by

\begin{equation}
	\phi_l(t)= \langle P_l(\hat{\mathbf{e}}_i(t_0) \cdot \hat{\mathbf{e}}_i(t_0+t)) \rangle\,.
	\label{reor}
\end{equation}

\section{R-simple systems and isomorphs}\label{sec:isom}

The virial $W$ quantifies the part of the pressure $p$ that derives from molecular interactions via the defining identity $pV=Nk_BT+W$. Liquids and solids may be classified according to the degree of correlation between the constant-volume thermal-equilibrium fluctuations of virial $W$ and potential energy $U$ \cite{ing12}. ``R-simple systems'' are those with strong $WU$ correlations; such systems are simple because their thermodynamic phase diagram is basically one-dimensional in regard to structure and dynamics \cite{ing12b,ing12,dyr14,dyr18a}.
The ``isomorph theory'' of R-simple systems was developed over the last decade \cite{I,IV}. 

The $WU$ Pearson correlation coefficient (which depends on the state point in question) is defined by

\begin{equation}
	R=\frac{\langle \Delta W \Delta U \rangle}{\sqrt{\langle (\Delta W)^2 \rangle \langle (\Delta U)^2 \rangle}}\,.
	\label{pearson}
\end{equation}
Here $\Delta$ gives the deviation from the equilibrium mean value. Many systems, including the standard Lennard-Jones and Yukawa fluids, have strong $WU$ correlations in their liquid and solid phases, whereas $R$ usually decreases significantly for densities below the critical density \cite{bel19a}. A system is considered to be R-simple whenever $R>0.9$ at the state points of interest \cite{I}. This is a pragmatic criterion, however, and, e.g., the simulations presented in this paper go below this value at high temperatures without significantly affecting the degree of isomorph invariance.

As mentioned, R-simple systems have curves in the phase diagram along which structure and dynamics are approximately invariant. These curves are termed \textit{isomorphs}. Isomorph invariance applies when data are presented in so-called reduced units. These units, which in contrast to ordinary units are state-point dependent, are given by letting the density $\rho$ define the length unit $l_0$, the temperature define the energy unit $e_0$, and density and thermal velocity define the time unit $t_0$,

\begin{equation}
	l_0=\rho^{-1/3},~~ e_0=k_{\rm B}T,~~ t_0=\rho^{-1/3}\sqrt{m/k_{\rm B}T}\,.
\end{equation}
Here $m$ is the molecule mass. Quantities made dimensionless by application of these units are termed ``reduced'' and marked with a tilde.

Strong virial potential-energy correlations arise whenever hidden scale invariance applies. This is the condition that the potential-energy ordering of same-density configurations is maintained under a uniform scaling of all coordinates \cite{sch14}. This is formally expressed as follows

\begin{equation}\label{HSI}
	U(\mathbf{R}_{\rm a})<U(\mathbf{R}_{\rm b})\Rightarrow U(\lambda \mathbf{R}_{\rm a})<U(\lambda \mathbf{R}_{\rm b})\,
\end{equation}
in which $\lambda$ is a scaling factor. Consider two configurations with the same potential energy, i.e., $U(\mathbf{R}_{\rm a})=U(\mathbf{R}_{\rm b})$. After a uniform scaling one has by \eq{HSI} $U(\lambda \mathbf{R}_{\rm a})=U(\lambda \mathbf{R}_{\rm b})$. By taking the derivative of this with respect to $\lambda$ one derives $W(\mathbf{R}_{\rm a})=W(\mathbf{R}_{\rm b})$ \cite{sch14}. Thus same potential energy implies same virial, resulting in a 100\% correlation between the $W$ and $U$ constant-volume fluctuations. For realistic systems \eq{HSI} is fulfilled only approximately, however, and in practice one rarely experiences perfect virial potential-energy correlations (this only applies when $U(\bR)$ is an Euler-homogeneous function).

Recall that a system's entropy $S$ is equal to that of an ideal gas at the same density and temperature plus an ``excess'' term deriving from the intermolecular interactions: $S=S_{\rm id}+S_{\rm ex}$. It can be shown that \eq{HSI} implies that the reduced structure and dynamics are invariant along the lines of constant excess entropy; these are by definition the system's isomorphs \cite{sch14}. The so-called density-scaling exponent $\gamma$ is defined by

 \begin{table}[!htbp]
	\centering
	\resizebox{0.24\textwidth}{!}{%
		\centering
		\begin{tabular}{c  c  c  c}
			\hline
			$\rho$ & $T$ & $R$ & $\gamma$ \\ [0.1ex] 
			\hline
			\hline
			0.400  & 0.400 & 0.956 & 9.46\\
            0.416  & 0.578 & 0.946 & 9.04\\	
            0.433  & 0.823 & 0.936 & 8.74\\		
            0.451  & 1.160 & 0.925 & 8.50\\
            0.469  & 1.619 & 0.905 & 8.28\\
            0.488  & 2.240 & 0.887 & 8.06\\
            0.508  & 3.079 & 0.868 & 7.92\\
 			0.529  & 4.211 & 0.854 & 7.85\\
 			0.550  & 5.770 & 0.854 & 8.00 \\              		
			\hline
		\end{tabular} 
	}
	\caption{Variation of density $\rho$, temperature $T$, virial potential-energy correlation coefficient $R$ (\eq{pearson}), and density-scaling exponent $\gamma$ (\eq{iso_gamma}) for nine state points on the isomorph generated from the reference state point $(\rho,T)=(0.4,0.4)$.}
	\label{table1}
\end{table}

\begin{equation}
	\gamma\equiv\left( \frac{\partial \ln T}{\partial \ln \rho} \right)_{S_{\rm ex}}=\frac{\langle \Delta W \Delta U \rangle}{\langle (\Delta U)^2 \rangle}\,.
	\label{iso_gamma} 	
\end{equation}
The second equality here is a general identity \cite{IV}, which is useful when the system is R-simple because \eq{iso_gamma} can then be applied for tracing out isomorphs without knowing the equation of state. For the simple Euler algorithm this is done by proceeding as follows. At a given state point $(\rho_1,T_1)$, by means of \eq{iso_gamma} one calculates $\gamma$ from the equilibrium fluctuations of potential energy and virial. From \eq{iso_gamma} one then predicts the temperature $T_2$ with the property that $(\rho_2,T_2)$ is on the same isomorph as $(\rho_1,T_1)$. If $\gamma=7$, for instance, for a one percent density increase a seven percent temperature increase will ensure that the new state point is on the same isomorph. In the simulations of this paper, however, in order to increase the accuracy of the generated isomorph, following Ref. \onlinecite{att21} we used instead the fourth-order Runge-Kutta algorithm for solving numerically \eq{iso_gamma} (involving density changes of order 1\% ). The resulting isomorph state points are given in Table I. 
{We note that the density-scaling exponent is generally significantly larger than for point-particle Lennard-Jones models where it is usually in the range 5-6. This must be a consequence of the spherical asymmetry because the same increase has been seen, e.g., for the asymmetric dumbbell and Lewis-Wahnstrom ortho-terphenyl models built of Lennard-Jones particles \cite{lew94,I,ing12b}. A quantitative explanation of this is missing, however, because a full isomorph theory of molecules is still not available.}

\begin{figure}[H]
	\centering
	\includegraphics[width=0.45\textwidth]{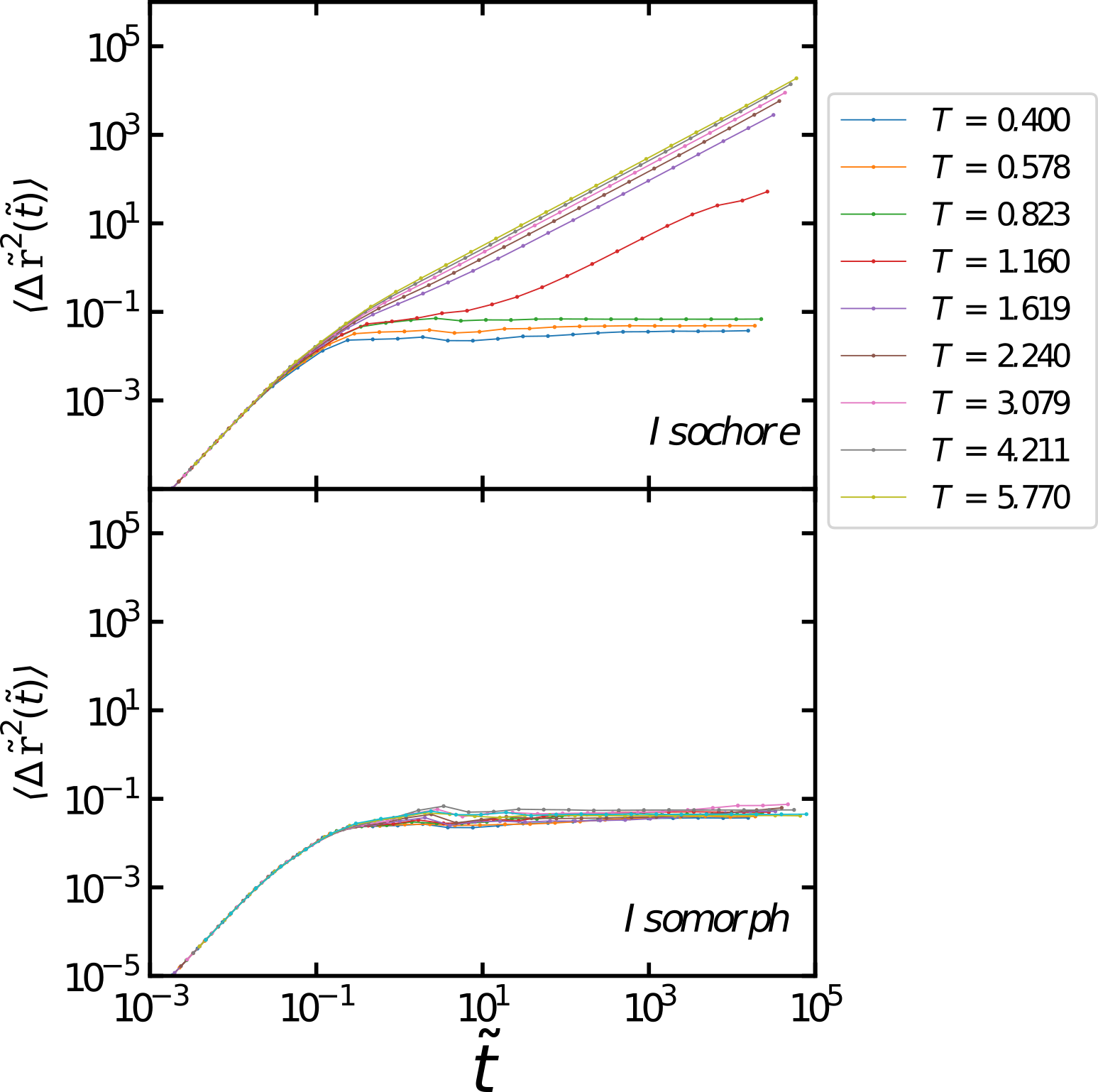}
	\caption{Reduced mean-square displacement as a function of reduced time along the $\rho=0.4$ isochore and along the isomorph generated from the reference state point $(\rho,T)=(0.4,0.4)$ (Table I).{ The colors here for the different temperatures are also used in Figs. 3-6.}}
	\label{fig2}
\end{figure}

\section{Structure and dynamics monitored along an isochore and an isomorph}

We begin the study by presenting results for the mean-square displacement as a function of time, which is predicted to be isomorph invariant in reduced units. \Fig{fig2} shows the results along the $\rho=0.4$ isochore (upper panel) and along the isomorph generated from the reference state point $(\rho,T)=(0.4,0.4)$ (lower panel), in both cases for the same nine temperatures. The isomorph data involve state points of more than a third density change and more than a factor of ten temperature change (Table I). {Note that the smectic B phase of the GB(5,3,2,1) model is found at higher densities than those of the isotropic and nematic phases studied in Ref. \onlinecite{meh22a}.}

The low-temperature state points along the isochore of \fig{fig2} are in the solid state as evident from the fact that the long-time mean-square displacement is constant. The high-temperature isochore state points, on the other hand, show diffusive long-time behavior and are consequently liquid. The fact that all mean-square displacement data collapse at short times in the ballistic regime for both the isochore and the isomorph is a straightforward consequence of the use of reduced units, because this leads to a reduced-unit thermal velocity that is the same at all state points. For the isomorph data, we see a fairly good collapse at all times, not just at short times. The minor deviations from perfect collapse are consistent with the fact that the virial potential-energy correlation coefficient $R$ is not very close to unity; in fact, $R$ goes below 0.9 at the four highest temperatures, compare Table I. {This feature might have to do with the short-time librational motion of the rods, which as shown below does not scale well in the isomorph sense.}

\begin{figure}[H]
    \centering
    \includegraphics[width=0.45\textwidth]{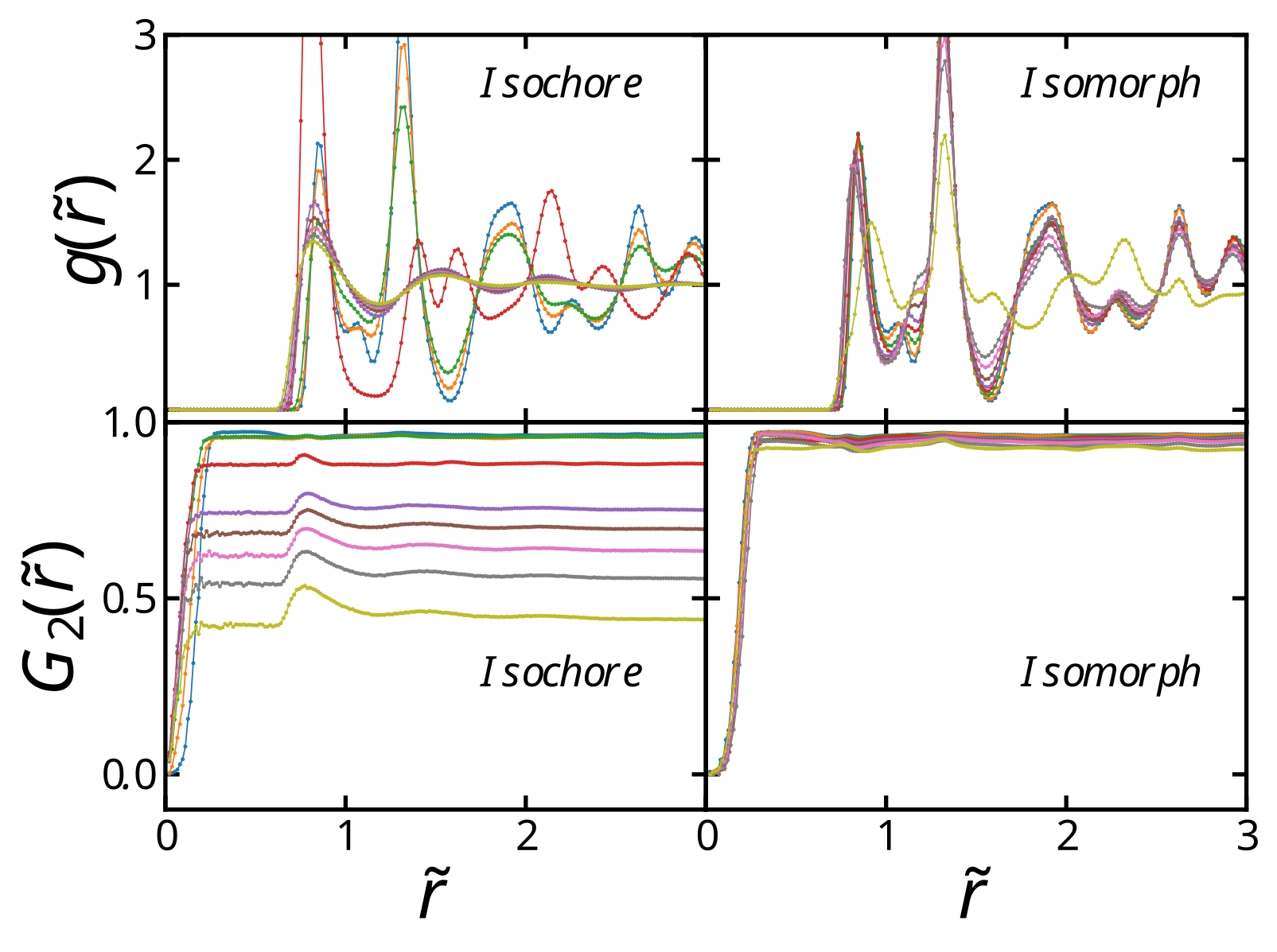}
    \caption{Structure along the isochore and the isomorph probed via the standard radial distribution function (upper panels) and the orientational radial distribution function defined in \eq{G_l} (lower panels), in both cases plotted as a function of the reduced pair distance $\tilde{r}$. {The colors used here and henceforth for the different temperatures are the same as those of Fig. 2.}}
    \label{fig3}
\end{figure}

\Fig{fig3} shows reduced-unit data for the radial distribution function $g(r)$ and the orientational radial distribution function $G_{2}(r)$ (\eq{G_l}) along the same isochore and isomorph. Figure \ref{fig3} shows no invariance along the isochore, but fair invariance along the isomorph. An exception to this is the highest temperature isomorph radial distribution function that deviates notably from the eight others. We have found that at this (and higher) temperatures, the smectic B phase undergoes a further transition {likely} involving a tilt of the average molecular orientation with respect to the smectic layers, similar to what has been reported by de Miguel \textit{et al.} \cite{de1991liquid}. Interestingly, this does not affect the isomorph invariance of other quantities than the radial distribution function, compare the $G_{2}(r)$ data of \fig{fig3}, as well as the data of later figures.

\begin{figure}[H]
    \centering
    \includegraphics[width=0.45\textwidth]{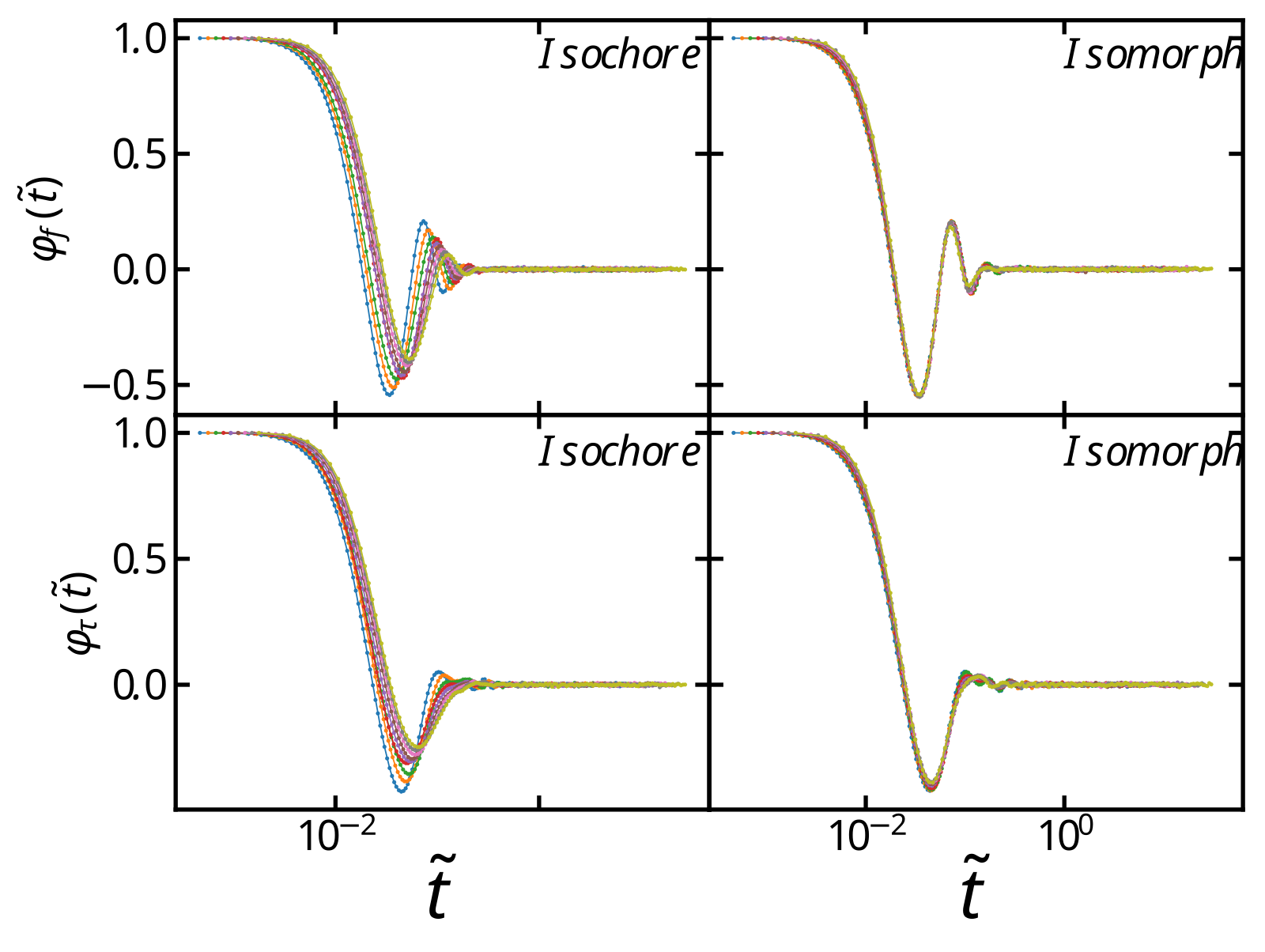}
    \caption{Normalized force (upper panels) and torque (lower panels) time-autocorrelation functions along the same isochore and isomorph as in the previous figures, plotted as functions of reduced time $\tilde{t}$.}
    \label{fig4} 
\end{figure}

{
Returning to dynamic properties, the normalized force and torque time-autocorrelation functions, i.e., the functions $\phi_A(t)/\phi_A(0)$
of \eq{autocorr} for $\bf{A}$ equal to the force and torque on the individual particles, respectively, are shown in Fig. \ref{fig4} as functions of the reduced time.}
Near-perfect scaling is observed for both functions along the isomorph, but not along the isochore.

\begin{figure}[H]
    \centering
    \includegraphics[width=0.45\textwidth]{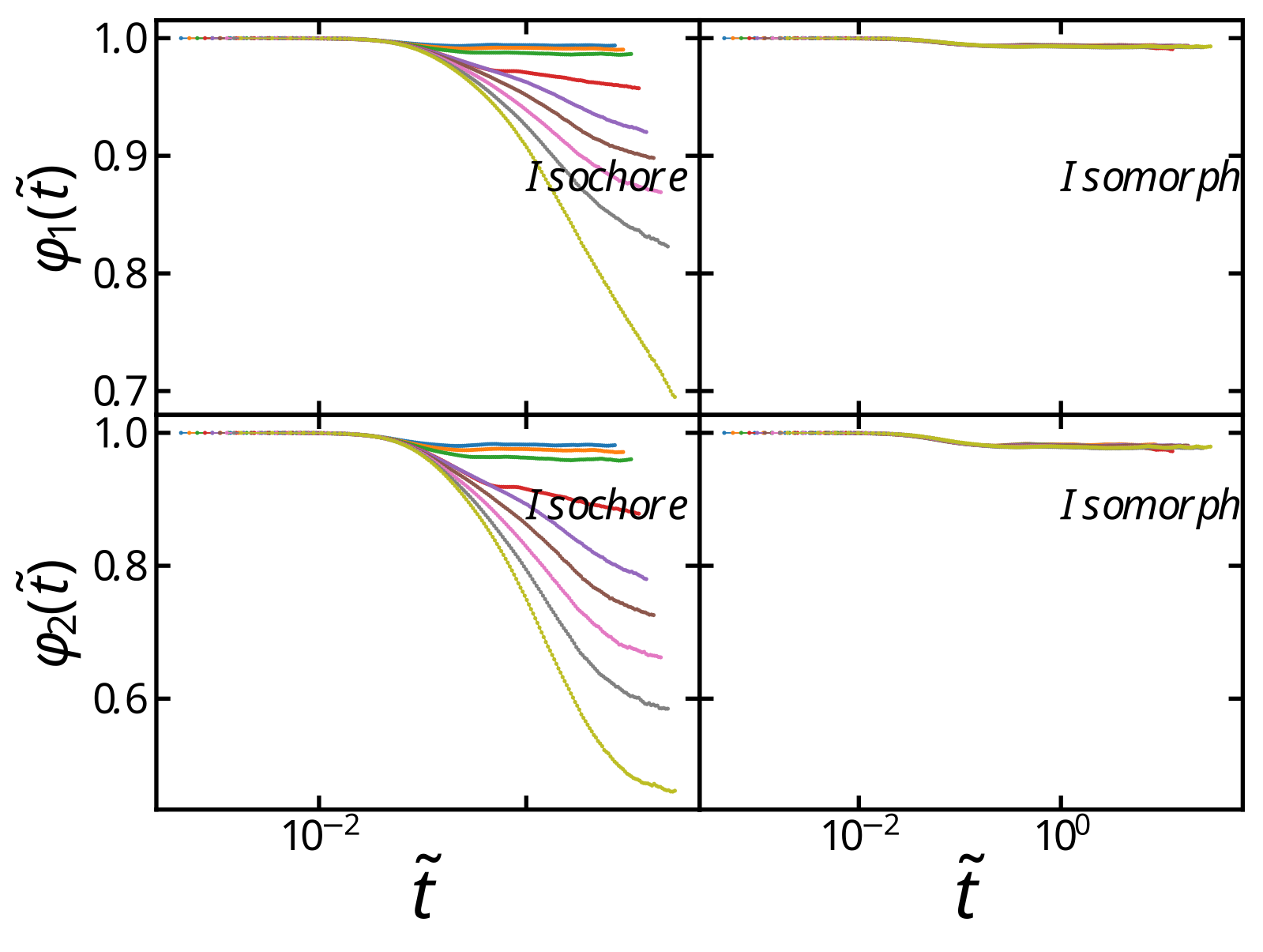}
     \caption{First- and second-order orientational order parameter time-autocorrelation functions along the isochore and isomorph, plotted as functions of reduced time. }
    \label{fig5}
\end{figure}

Figure \ref{fig5} shows the first- and second-order orientational time-autocorrelation functions along the isochore and the isomorph. These functions both decay to zero at the highest density studied on the isochore, which is not the case for the isomorph along which invariant dynamics is observed.

\begin{figure}[H]
    \centering
    \includegraphics[width=0.45\textwidth]{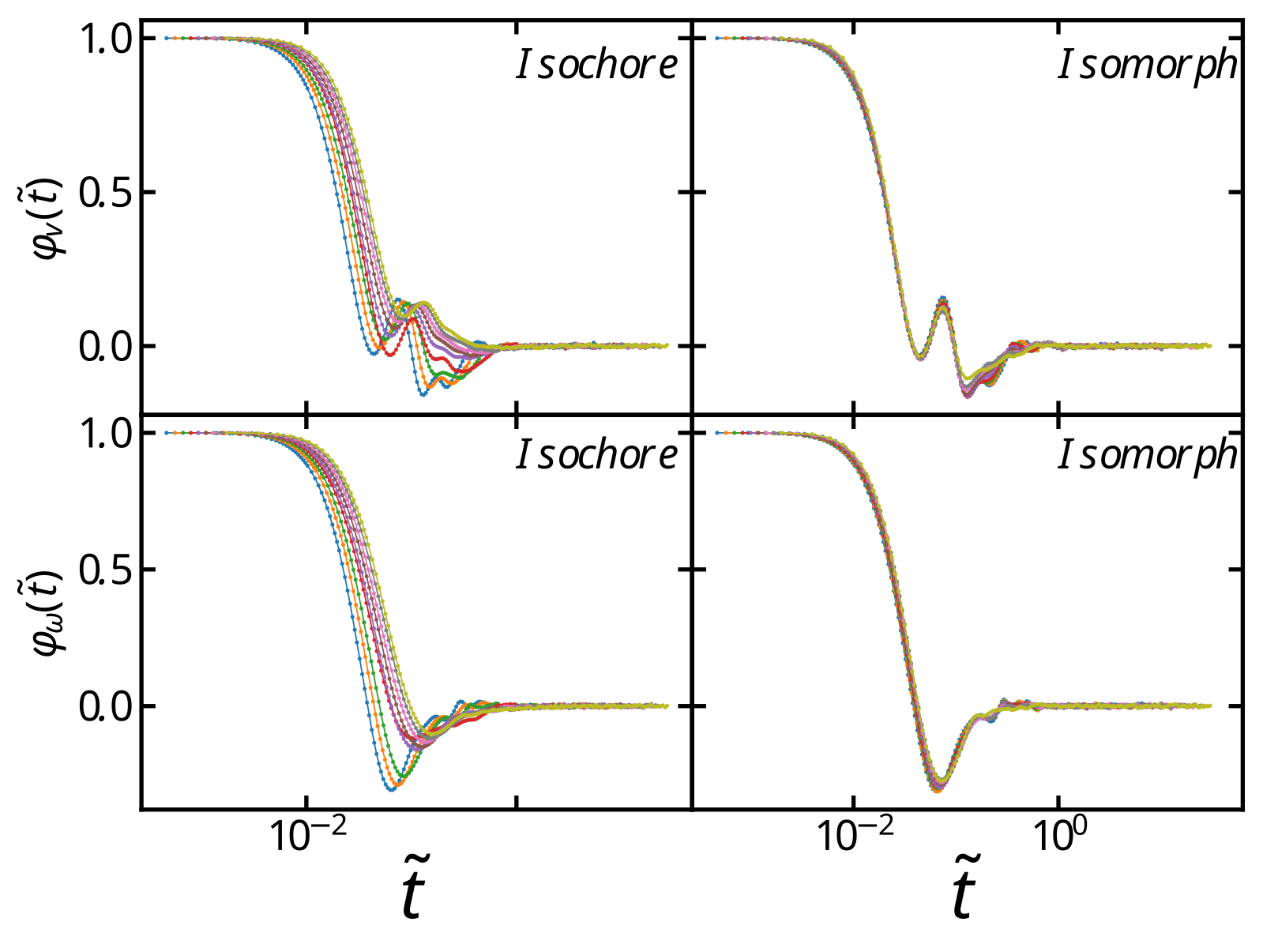}
     \caption{Normalized velocity and angular velocity time-autocorrelation functions along the isochore and isomorph, plotted as functions of reduced time.}
    \label{fig6}
\end{figure}

We finish the study by showing the normalized velocity and angular velocity time-autocorrelation functions in Fig. \ref{fig6}. Again, good isomorph invariance is observed at all times, though with minor deviation at intermediate times for the velocity time-autocorrelation function.

\section{Summary}

We have shown that the isomorph theory can be used to understand GB liquid crystals in the smectic B phase, because the thermodynamic phase diagram is here effectively one-dimensional in the sense that the reduced-unit structure and dynamics are approximately invariant along the isomorphs. Our previous paper \cite{meh22a} showed that the same applies for the isotropic and nematic phases of the GB(3,5,2,1) model. This means that most of the GB(3,5,2,1) phase diagram is effectively one-dimensional in regard to structure and dynamics. We note that this property is not limited to a particular GB model; thus an earlier publication demonstrated the existence of isomorphs in the GB(0.345,0.2,1,2) model that forms a discotic liquid-crystal phase at low temperatures \cite{meh22}. 
{-- The GB potential is unique in the field of liquid-crystal models in that through a gradual reduction of the parameters $\chi$ and $\chi'$ of \eq{GB_sigma} and \eq{GB_epsilon}, the Lennard-Jones potential is recovered. It is an interesting question whether one would find isomorph invariance behavior in other models of rods, such as a rigid line of Lennard-Jones interaction centers.}

We demonstrated above that the GB(3,5,2,1) model exhibits good invariance of the reduced-unit structure and dynamics along the studied isomorph. In conjunction with our previous study \cite{meh22a}, the existence of isomorphs in the GB model can now be used to explain the observed behavior of liquid crystals, for instance the so-called density scaling, which is the fact that the reduced dynamics is invariant along lines of constant $\rho^\gamma/T$ \cite{rol05,satoh2006characteristic}. Studies remain to investigate whether other smectic phases of the GB model also exhibit strong virial potential-energy correlations and thus the existence of isomorphs. 
{It would be interesting, in particular, to investigate the effect of varying the moment of inertia, given the fact that the fixing of this quantity upon a density change formally violates isomorph invariance of the dynamics, but was found above to have little effect in practice.} Also, it would be interesting to investigate systematically the vast parameter space of the GB potential from the hidden-scale-invariance perspective.

\begin{acknowledgments}
	This work was supported by the VILLUM Foundation's \textit{Matter} grant (16515).
\end{acknowledgments}

%


\end{document}